\documentclass[aps, reprint, nofootinbib, amsmath,amssymb, aps, prd]{revtex4-2}

\usepackage{graphicx}
\usepackage{dcolumn}
\usepackage{bm}
\usepackage{amsfonts}
\usepackage{amsmath}
\usepackage{amssymb}
\usepackage{mathrsfs}
\usepackage{graphicx}
\usepackage{cancel}
\usepackage{bbm}
\usepackage{bm}
\usepackage{mathtools}
\usepackage{physics}
\usepackage{tensor}
\usepackage{array}
\usepackage{float}
\usepackage{wrapfig}
\usepackage{soul}
\usepackage{xcolor}
\usepackage{url}
\usepackage{tabu}
\usepackage{tikz}
\usepackage{hyperref}
\usepackage{cleveref}

\newcommand{\ebd}{\vcentcolon =}

\newcommand{\upsigma}{\ket{\uparrow_\sigma}}

\newcommand{\downsigma}{\ket{\downarrow_\sigma}}

\newcommand{\up}{\ket{\uparrow}}

\newcommand{\down}{\ket{\downarrow}}

\hypersetup{
colorlinks=true,
linkcolor=blue,
citecolor=red,
urlcolor=cyan
}

\begin{document}

\preprint{APS/123-QED}

\title{Indefinite probabilities in quantum spacetime:\\ A deepening of unpredictability}

\author{Vittorio D'Esposito}
\thanks{\href{mailto:vittorio.d.esposito@uniroma2.it}{vittorio.d.esposito@uniroma2.it}}

\affiliation{%
 Dipartimento di Fisica, Universit\`a degli studi di Roma Tor Vergata
}%
\affiliation{INFN, Sezione di Roma Tor Vergata, 00133 Roma, Italy}

\author{Giuseppe Fabiano}%
\thanks{\href{mailto:gfabiano@lbl.gov}{gfabiano@lbl.gov}}
 \affiliation{%
Physics Division, Lawrence Berkeley National Laboratory,
}%
\affiliation{Centro Ricerche Enrico Fermi—Museo Storico della Fisica e Centro Studi e Ricerche “Enrico Fermi”, Roma, Italy}

\author{Domenico Frattulillo}
\thanks{\href{mailto:domenico.frattulillo@unina.it}{domenico.frattulillo@unina.it}}

\affiliation{
 Dipartimento di Fisica ``E. Pancini'', Università di Napoli Federico II,
}%
\affiliation{
INFN, Sezione di Napoli, 80126 Napoli, Italy
}%

 \affiliation{%
Institute for Quantum Optics and Quantum Information - Vienna, Austrian Academy
of Sciences, Vienna, Austria
}%

\begin{abstract}
Non-commutative spacetime and quantum groups have been argued to capture non-classical features of spacetime and its symmetries in quantum gravity. In this letter, we show that employing the $SU_q(2)$ quantum group to describe rotational symmetry for spin-$\frac{1}{2}$ systems and Stern-Gerlach apparatuses leads to the description of probabilities of outcomes of spin measurements in terms of non-commuting operators. 
As a result, we obtain an uncertainty principle between different probability operators, realising a notion of indefinite probabilities.
This is then reflected in the non-commutativity of the entries of the rotation matrix relating the reference frames of two observers, hence fundamentally preventing them from sharply measuring their relative orientation.
\end{abstract}

\maketitle

\section{Introduction}\label{Introduction}

Investigations at the interface between quantum theory and gravity must reckon with a foundational aspect of general relativity: diffeomorphism invariance implies that only relational quantities carry physical content. This means, for instance, that the reference frames used by observers to describe events have no physical meaning, and only the relations between these frames do. It is therefore essential to investigate the properties of such relational quantities in models of quantum gravity. As an example, consider two observers, Alice and Bob, whose reference frames differ only by a rotation. The only physical information resides in their relative orientation in space. If the latter is unknown to them, can they establish a communication protocol to measure this relational quantity? In classical spacetime, one possibility is for Alice to send Bob spins aligned with the axes of her reference frame. Bob measures these spins with Stern-Gerlach devices aligned along his own reference frame, and can reconstruct the rotation matrix connecting the two frames by means of the relation \cite{Bartlett:2006tzx, Amelino-Camelia:2022dsj, DEsposito:2024wru}
\begin{equation}\label{rotation matrix classical}
    R_{ij} = \mel{\uparrow_{i}^A}{\sigma_{j}^{B}}{\uparrow_{i}^A} = 2 \,p_{ij} - 1\;,
\end{equation}
where $i,j=x,y,z$ and $p_{ij}$ is the probability that Bob obtains spin up as a result of a measurement performed with his device aligned with his axis $j$ on Alice's spins aligned with her axis $i$.

What changes might we expect for this protocol once quantum gravity effects are taken into account? In the quantum gravity regime, geometric quantities are expected to acquire non-classical features, such as discrete spectra or non-commutativity \cite{Rovelli:1994ge, Ashtekar:1996eg, Freidel:2002hx, Amelino-Camelia:2008fcv, Majid:2006xn, Lizzi:2018qaf, Majid:2026ptz}. Applying this general expectation to the example above, where a relational geometrical quantity is inferred from probabilities of measurement outcomes, the non-classicality of the former should be reflected in non-classical features of the latter. These considerations suggest that, in a non-classical spacetime, probabilities themselves could become indefinite, in a genuinely quantum sense. Such a notion of \emph{indefinite probabilities} would deepen the fundamental unpredictability of quantum theory, extending it beyond measurement outcomes to the probabilities of those outcomes themselves.

We set out to explore this largely overlooked aspect of quantum spacetime research within the framework of non-commutative geometry \cite{Snyder:1947,Majid:2006xn}, a leading candidate for the effective description of spacetime emerging from  quantum gravity at length scales above the Planck length \cite{Seiberg:1999vs, Douglas:2001ba, Brandenberger:2002nq, Freidel:2005me, Oriti:2009pb, Amelino-Camelia:2016gfx,Szabo:2001kg, Amelino-Camelia:2008aez}. The idea is that spacetime coordinates are promoted to non-commuting operators. To preserve the relativistic description of physical laws in this setting \cite{Lizzi:2018qaf}, the standard Lie groups of spacetime symmetries are replaced by quantum groups \cite{majid_1995}, where numerical group parameters are promoted to non-commutative operators themselves. In recent years, there has been a growing interest \cite{DEsposito:2024wru, Arzano:2022nlo, Ballesteros:2025ypr, Gubitosi:2021itz,Arzano:2026gng,Arzano:2026fbz}, in studying quantum mechanical systems with quantum group symmetries, and in \cite{Ballesteros:2025ypr} it has also been shown that a quantum deformation of the Galilei group describes the structure of quantum reference frame transformations for Galilean systems introduced in \cite{Giacomini:2017zju}.

In this letter, we introduce the notion of indefinite probabilities in quantum spacetime by promoting the rotation symmetry group $SU(2)$ to its quantum deformation $SU_q(2)$ \cite{Woronowicz:1987vs}. This quantum group appears in different contexts in quantum gravity research \cite{Major:1995yz, Freidel:1998pt, Girelli:2022foc, Bianchi:2011uq}, where the parameter $q$ is considered to be a function of the Planck length and the cosmological constant. We build on previous work \cite{DEsposito:2024wru}, in which we introduced the framework of \textit{doubly quantum mechanics} (DQM), a deformation of quantum theory in which not only the phase space of physical systems, but also their geometrical configurations, are described by quantum degrees of freedom. Focusing on the spin sector of the theory, we showed that promoting rotational symmetry to $SU_q(2)$ naturally leads to the conclusion that probability itself becomes a quantum mechanical operator, acting on the geometry states that describe the orientations of spin beams and Stern-Gerlach apparatuses. In general, repeated inferences of the same probability under identical experimental conditions are therefore distributed with a non-vanishing variance that depends on the geometry state of the configuration. In the present work, we introduce a mechanism, known as \emph{braiding}, that guarantees the full covariance of the framework by imposing non-trivial commutation relations between operators associated with different spin beams or Stern-Gerlach apparatuses. As a result, probability operators associated with different measurements no longer commute, thus endowing them with genuinely quantum features: measuring one probability prevents one from predicting with certainty the outcome of a subsequent measurement of another probability, thus realising a notion of \textit{indefinite probabilities}. This represents a new feature of quantum theory, arising from embedding quantum spacetime effects into its very foundations. It is therefore impossible, to sharply measure the relative orientation between two observers with the protocol described by \eqref{rotation matrix classical}, even in the ideal case of infinite precision and infinite resources, since different elements of the rotation matrix do not commute. This means that the relative orientation between the two observers becomes a genuinely quantum object.

The paper is organised as follows. In \cref{sec: Review and braiding} we briefly review the doubly quantum mechanics framework. In \cref{sec: braided DQM} we employ the braiding formalism to guarantee the covariance of our framework and introduce the notion of probability measurement apparatuses. In \cref{sec: indefinite probabilities} we derive the main result of the paper, \emph{i.e.} the non-commutativity of probabilities that allows us to introduce an uncertainty principle between these observables. In turn, this translates into an uncertainty principle for the elements of the rotation matrix relating two observers (cf. \cref{rotation matrix classical}). Finally, in \cref{sec: discussion} we conclude by summarising our results and discussing some potential connections with the quantum reference frames research programme \cite{Giacomini:2017zju, hoehn2019trinity,giacomini2019relativistic, streiter2020relativistic, de2020quantum, Mikusch:2021kro, hoehn2021quantum, de2021entanglement, giacomini2021spacetime, Cepollaro2024, Ballesteros:2025ypr, Garmier2025}.

\section{The principles of DQM}\label{sec: Review and braiding}

We begin by recalling the basic constructions of spin states and Pauli matrices in standard quantum theory, which we will generalise to the non-commutative case. With respect to some reference frame $A$, a generic spin state $\ket{\psi^A}$ and Pauli matrix $\sigma^A$ are uniquely identified by two distinct elements of $SU(2)$, which act on a reference state and Pauli matrix aligned along some axis. Concretely, a possible choice is $\ket{\psi^A}=U_s \ket{\uparrow^A}$ and $\sigma^A = U_a \sigma^A_z U_a^\dagger$, where $U_s,U_a$ are $SU(2)$ matrices and $\sigma_z^A\ket{\uparrow^A}=\ket{\uparrow^A}$. To describe these same quantities with respect to another reference frame, $B$, we apply an $SU(2)$ transformation $U_r$: $\ket{\psi^B}=U_r\ket{\psi^A}$ and $\sigma^B=U_r\sigma^AU_r^\dagger$.

The construction above is readily generalised when the $SU(2)$ group is deformed into the $SU_q(2)$ quantum group \cite{DEsposito:2024wru}. This quantum group is defined by deforming the algebra of functions on $SU(2)$ in a non-commutative way. Its spin-$\frac{1}{2}$ representation is given by the matrix \cite{Woronowicz:1987vs}
\begin{equation}
\label{eq:suq2mat}
    U^q=
\begin{pmatrix}
    \alpha & -q\gamma^*\\
    \gamma & \alpha^* 
\end{pmatrix}\;,
\end{equation}
where $*$ denotes Hermitian conjugation and $\alpha,\gamma$ are non-commutative functions on the group manifold satisfying the commutation relations
\begin{equation}\label{suq2 algebra}
\begin{gathered}
     \alpha \gamma = q\, \gamma \alpha\;,\;\; 
\alpha \gamma^*= q\,  \gamma^*\alpha 
\;,\;\; \gamma\gamma^*=\gamma^*\gamma 
\\ \gamma^*\gamma+\alpha^*\alpha = \mathbbm{1}  \;,\;\;  \alpha \alpha^*-\alpha^*\alpha=(1-q^2)\gamma^*\gamma\;.
\end{gathered}
\end{equation}
The definitions above hold for a generic $q\in \mathbb{C}$, although in our framework we focus on the case $0<q\leq 1$. In the quantum gravitational setting it is expected to be very close to $1$, \emph{e.g.} $q = e^{\ell_P\sqrt{\abs{\Lambda}}}$ in \cite{Girelli:2022foc}. In the limit $q\to1$ the classical $SU(2)$ group is recovered. 

In this framework, physical objects of interest for an observer, Alice, can be constructed in terms of $q$-spinors \cite{Xing-ChangSong_1992,DEsposito:2024wru}, which are $SU_q(2)$ operator-valued vectors. Consider two distinct copies of $SU_q(2)$ denoted by $U_s^q$ and $U_a^q$, which are associated to the spin and Stern-Gerlach directions, respectively. Their spin-$\frac{1}{2}$ representations can be written in terms of operators $(x,y)$ and $(a,c)$, respectively, satisfying commutation relations analogous to those of $(\alpha, \gamma)$ in \eqref{eq:suq2mat}. These operators act on the Hilbert space, $\mathcal{H}^A=\mathcal{H}^q_s\otimes \mathcal{H}^q_a$, of geometry states, which encode information on the direction of the spin system and the Stern-Gerlach apparatus. We will comment more on the properties of this Hilbert space below.

The $q$-spinors associated to the two copies are then defined as
\begin{equation}
\label{eq:q-spinors_Alice}
\begin{aligned}
    &u^\alpha \coloneqq U^q_s\up=\begin{pmatrix}
           x\\
           y
         \end{pmatrix}\,,\, \bar{u}_\alpha \coloneqq\bra{\uparrow}(U^q)^\dagger = (x^*,y^*)\,,\\
         &w^\alpha \coloneqq U^q_a\up=\begin{pmatrix}
           a\\
           c
         \end{pmatrix}\,,\, \bar{w}_\alpha \coloneqq\bra{\uparrow}(U^q)^\dagger = (a^*,c^*)\,.
\end{aligned}
\end{equation}
where $\up=(1,0)$ and $\down=(0,1)$ in the $z$-basis.
Here, indices are raised and lowered as
$u^\alpha = u_\beta\,\epsilon^{\beta\alpha}$ and $u_\alpha = u^\beta\,\epsilon_{\beta\alpha}$, and similarly for $w$, where $\epsilon$ denotes the $q$-deformation of the Levi-Civita symbol
\begin{equation}
    \epsilon^{\alpha\beta} = \begin{pmatrix}
    0 & 1\\
    -q & 0 
\end{pmatrix}\;,\;\; \epsilon_{\alpha\beta} = \begin{pmatrix}
    0 & -q^{-1}\\
    1 & 0 
\end{pmatrix}\;.
\end{equation}

The generic spin up ket is then simply $\ket{\psi}=u^\alpha$ while its bra is $\bra{\psi}=\bar u_{\alpha}$. The generic Pauli matrix $\sigma^q$ is defined by acting with $U^q_a$ on $\sigma^q_z=\mathrm{diag}(q,-q^{-1})$ as $\sigma^q\coloneq U^q_{a} \,\sigma_z^q\, {U^q_{a}}^\dagger$. It can also be written in terms of projectors as
\begin{equation}
    \sigma^q = q\, \Pi_{\uparrow_\sigma} -q^{-1}\, \Pi_{\downarrow_\sigma}\;,\label{generic_Pauli}
\end{equation}
where $\upsigma=w^\alpha$ and $\downsigma=\bar w^\alpha$. 
Aside from some factors of $q$, these expressions resemble those of $SU(2)$, with the difference that the commutative factors of $SU(2)$ are now non-commutative operators.

From these constructions, it follows quite naturally that probabilities are promoted to operators via
\begin{equation}\label{probabilities definition}
    P\qty(\uparrow_\sigma) \ebd \mel{\psi}{\Pi_{\uparrow_\sigma}}{\psi}\;,\;\;P\qty(\downarrow_\sigma)\ebd \mel{\psi}{\Pi_{\downarrow_\sigma}}{\psi}\;,
\end{equation}
and satisfy the following properties
\begin{equation}\label{probabilities properties}
    P\qty(\uparrow_\sigma) \geq 0\,,\,P\qty(\uparrow_\sigma)^\dagger = P\qty(\uparrow_\sigma)\,,\, P\qty(\uparrow_\sigma) + P\qty(\downarrow_\sigma) = \mathbbm{1}\;.
\end{equation}

According to \eqref{probabilities properties}, probabilities of observing given outcomes are described by positive semi-definite self-adjoint operators on $\mathcal{H}^A$ and not by non-negative real numbers as in standard quantum mechanics. In general, geometry states are written in terms of superpositions of probability eigenstates \cite{DEsposito:2024wru} and thus probabilities of observing given outcomes will be themselves characterised by probability distributions with non-zero variance. 
In the DQM framework, probability is thus promoted to a quantum observable of the theory, whose quantum mechanical features are encoded in geometry states. By definition,  an observable must be associated with a measurement procedure. In this context of spin measurements, the measurement apparatus consists of a Stern-Gerlach device and a beam of spin-$\frac{1}{2}$ particles. The output of this measurement device is a number between 0 and 1 related to the fraction of spin up (down) particles in the beam. In a real-world experiment, this device will be equipped with a finite number $N$ of particles, and the resolution of the probability measurement will increase with $N$. In DQM, as a consequence of our non-commutativity effects, for a sufficiently high value of $N$, two probability measurements performed in identical geometrical configurations may yield incompatible outcomes within their experimental uncertainty.

\section{Braided DQM}\label{sec: braided DQM}

We now turn to an aspect that was not addressed in our exploratory work \cite{DEsposito:2024wru}. Up to this point, we have left the cross-commutation relations between different copies of $SU_q(2)$ unspecified, and we will now show that a non-trivial choice of such relations is in fact necessary to ensure the full covariance of our framework. Concretely, if operators belonging to different copies of $SU_q(2)$ are taken to commute, they will generically fail to do so once rotated by yet another copy of $SU_q(2)$. The prescription that fixes the cross-commutation relations between different copies of a quantum group is known as \textit{braiding} \cite{majid_1995}, and it guarantees that the defining relations of the $SU_q(2)$ operators take the same form in every reference frame. While the general construction of \cite{DEsposito:2024wru} remains applicable and can still be used to define a quantum theory of spins on a quantum spacetime, with several of our previous results surviving qualitatively unchanged, the braiding formalism reveals a substantially richer underlying structure.

\subsection{The braiding construction}\label{subsec: braiding}

Consider the scenario in which Alice wants to rotate her lab equipment (spins and Stern-Gerlach apparatus). Following \cite{DEsposito:2024wru}, we introduce a further copy of $SU_q(2)$, denoted $U^q_r$, which implements this symmetry transformation. The $q$-spinors describing the orientations of the spin states and the Stern-Gerlach apparatus transform as
\begin{equation}
    u' = U^q_r\, u \;, \; w' = U^q_r\, w \,,
\end{equation}
and the probability \eqref{probabilities definition} is invariant under this transformation, as expected in a covariant framework. However, assuming that the generators of the two copies commute, \textit{i.e.}, $uw = wu$, leads to an inconsistency: this relation is not preserved under $SU_q(2)$ transformations, namely $u'w' \neq w'u'$.

A standard way out of this impasse in non-commutative geometry is the introduction of so-called \textit{braiding} commutation relations \cite{majid_1995}. The general idea is that when multiple distinct copies of non-commutative coordinate functions are transformed under the same quantum group symmetry, the non-commutative nature of the transformation requires non-trivial commutation relations between the copies in order to preserve covariance. In the specific case of $SU_q(2)$, these relations are well known \cite{Xing-ChangSong_1992} and take the form
\begin{equation}\label{braiding equations}
\begin{gathered}
    u^\alpha \,w^\beta = q^{-1} R^{\alpha\beta}_{\gamma\delta}\, w^\gamma \, u^\delta\;, \\
    u^\alpha\, \bar{w}^\beta = R^{\alpha\beta}_{\gamma\delta}\, \bar{w}^\gamma \, u^\delta\;.
\end{gathered}
\end{equation}
where
\begin{equation}
    R^{\alpha\beta}_{\gamma\delta} = q\,\delta^\alpha_\gamma\,\delta^\beta_\delta + \epsilon^{\alpha\beta}\epsilon_{\gamma\delta}\;,
\end{equation}
is the so-called $R$-matrix. By explicit computation, it can be shown that $q$-spinors $u',w'$ close the commutation relations \eqref{braiding equations}, thus preserving covariance. For an arbitrary number of $q$-spinors, braiding relations of the same form as \eqref{braiding equations} must be imposed, once an ordering prescription for the $q$-spinors is fixed. As a side remark, we note that the braiding relations \eqref{braiding equations} imply that the pairs of operators describing the spin system and the Stern-Gerlach apparatus do not act locally on Hilbert spaces $\mathcal{H}^q_s$ and $\mathcal{H}^q_a$, but on the Hilbert space $\mathcal{H}^A$ as a whole. The explicit form of the action of the relevant $SU_q(2)$ operators on $\mathcal{H}^A$ is not essential for the main results of our paper. We plan to investigate this in future studies.

\subsection{Probability measurement apparatuses}

The discussion above can be generalised for an observer performing an arbitrary number of different probability measurements. Since each probability measurement involves a beam of spin-$\frac{1}{2}$ particles and a Stern-Gerlach apparatus, whose orientations are described by distinct copies of $SU_q(2)$, it is convenient to define a \emph{probability measurement apparatus} by the set
\begin{equation}\label{probability measurement apparatus}
    \widehat{\mathcal{M}}_i \ebd \Big\{u^\alpha_{i}, \bar{u}^\alpha_{i}, w^\alpha_{i}, \bar{w}^\alpha_{i}\Big\}\;,
\end{equation}
where the spinors are defined as in \eqref{eq:q-spinors_Alice}, with $SU_q(2)$ operators carrying subscripts $i$. These define the directions of the spins in the beam and the Stern-Gerlach, respectively. With this definition, the probabilities of observing spin up and down are still given by \eqref{probabilities definition} and satisfy the properties \eqref{probabilities properties}. They are written explicitly as 
\begin{equation}\label{probability operators braided}
\begin{gathered}
    P_i\qty(\uparrow) = \bar{u}^{i}_\alpha\, w^\alpha_{i} \,\bar{w}^{i}_\beta \,u^\beta_{i}\;,  \\
    P_i\qty(\downarrow) = \mathbbm{1} - P_i\qty(\uparrow)\;.
\end{gathered}
\end{equation}

An observer equipped with a set of $n$ probability measurement apparatuses $\big\{\widehat{\mathcal{M}}_i\big\}_{i=1}^n$ is described by means of $n$ $q$-spinors associated to spin-$\frac{1}{2}$ beams and $n$ $q$-spinors associated to Stern-Gerlach apparatuses. As anticipated in the preceding section, we require braiding relations analogous to \eqref{braiding equations} for each pair of these $q$-spinors, so we have to fix an ordering prescription. To do so, we group all the copies of the $SU_q(2)$ algebra as $(S_1,\ldots,S_n,SG_1,\ldots, SG_n)$, where $S_i$ and $SG_j$ denote the copies of the algebra relative to the $i$-th spin beam and $j$-th Stern-Gerlach apparatus. The following braiding commutation relations are then imposed

\begin{equation}\label{braiding euqations n copies}
\begin{gathered}
   X^\alpha_{i} \,X^\beta_{j} = \frac{1}{q} R^{\alpha\beta}_{\gamma\delta}\, X^\gamma_{j} \,X^\delta_{i} \\
    X^\alpha_{i}\, \overline{X}^\beta_{j} = R^{\alpha\beta}_{\gamma\delta}\, \overline{X}^\gamma_{j} \, X^\delta_{i}
\end{gathered}\;,\, i \leq j\,,\, i,j=1,\ldots, 2n\,,
\end{equation}
where $X_i^\alpha = u_i^\alpha$ for $i=1,\ldots,n$ and $X_i^\alpha = w_i^\alpha$ for $i=n+1,\ldots,2n$, and analogously for $\overline{X}^\gamma_{i}$. This ordering is such that all spin beams satisfy the same form of braiding relations with all the Stern-Gerlach devices. With these relations, it is clear that the $q$-spinors defining different probability measurements do not commute, and this leads to non-commuting, probability operators.
\vspace{15pt}
\section{Indefinite probabilities}\label{sec: indefinite probabilities}
We are now in a position to discuss the striking feature introduced by the braided structure. A lengthy but straightforward calculation shows that two probability operators $P_i(\uparrow)$ and $P_j(\uparrow)$, relative to two different probability measurements, do not commute. The result to all orders in $q$ is reported in Appendix \ref{appendix: computations}. Here, we limit our analysis to the leading order expansion in $1-q$, both to gain physical insight into the meaning of this commutator and in view of the expected smallness of the quantum deformation scale $1-q$. By doing so, the commutator reduces to
\begin{equation}\label{commutators probabilities}
    \begin{gathered}
    \Big[P_i\qty(\uparrow)\,,\,P_j\qty(\uparrow)\Big] = \frac{i}{2}(1-q)\Big\{\qty( \widehat{\mathbf{m}}_i+ \widehat{\mathbf{m}}_j)\cdot \qty( \widehat{\mathbf{n}}_i\times  \widehat{\mathbf{n}}_j)\\ - \qty( \widehat{\mathbf{n}}_i+ \widehat{\mathbf{n}}_j)\cdot \qty( \widehat{\mathbf{m}}_i\times  \widehat{\mathbf{m}}_j)\Big\}\,,
    \end{gathered}
\end{equation}
where the dot and cross denote scalar and vector products, and we have defined the \emph{quantum direction operators} at zero order in $(1-q)$ as
\begin{equation}\label{directions}
    \begin{gathered}
        \widehat{\mathbf{n}}_i \ebd \qty(x_i\, y^*_i + y_i\, x^*_i\,,\, \mathbbm{i}(y_i\, x_i^* - x_i\, y_i^*)\,,\, 1 - 2\,y_i \,y_i^*)\,,\\
        \widehat{\mathbf{m}}_i \ebd \qty(a_i\, c^*_i + c_i\, a^*_i\,,\, \mathbbm{i}(c_i\, a_i^* - a_i\, c_i^*)\,,\, 1 - 2\,c_i \,c_i^*)\,,
    \end{gathered}
\end{equation}
which coincide with their classical definitions in the commutative limit. These quantum direction operators can be taken to be commutative in \eqref{commutators probabilities}, since every contribution coming from non-commutativity would lead to sub-leading $(1-q)^2$ terms. Their common eigenstates are direction eigenstates $\ket{\theta,\omega}$, with $\theta\in [0,\pi]$ and $\omega\in [0,2\pi]$ denoting a classical direction in space. In this approximation, the action of a generic quantum direction operator $ \hat {\mathbf n}$ on $\ket{\theta,\omega}$ is 
\begin{equation}\label{direction operators action}
     \widehat{\mathbf{n}} \ket{\theta,\omega} = \qty(\sin \theta \cos \omega \,,\; \sin \theta \sin \omega\,,\;\cos \theta)\ket{\theta,\omega}\;,
\end{equation}
\emph{i.e.}, the eigenvalues of a direction operator are the classical directions parametrised by the angles $(\theta,\omega)$.

With the commutator in \eqref{commutators probabilities} we can easily derive an uncertainty principle for probability measurements. Indeed, we have 
\begin{widetext}
    \begin{equation}\label{uncertainty principle probabilities}
    \Delta_\Phi P_i(\uparrow) \Delta_\Phi P_j (\uparrow) \geq \frac{(1-q)}{4} \abs{\expval{\qty( \widehat{\mathbf{m}}_i+ \widehat{\mathbf{m}}_j)\cdot \qty( \widehat{\mathbf{n}}_i\times  \widehat{\mathbf{n}}_j) - \qty( \widehat{\mathbf{n}}_i+ \widehat{\mathbf{n}}_j)\cdot \qty( \widehat{\mathbf{m}}_i\times  \widehat{\mathbf{m}}_j)}_{\Phi}}\;,
\end{equation}
\end{widetext}
where $\expval{\cdot}_\Phi = \mel{\Phi}{\cdot}{\Phi}$ denotes the expectation value, and $\ket{\Phi}$ is the state describing the geometry of the experimental setups, \emph{i.e.} the directions of the spin systems and the Stern-Gerlach apparatuses used to perform the measurements. In general, these states could also describe scenarios in which we have superposition and entanglement of classical directions \eqref{direction operators action}. In the simple case where $\ket{\Phi}$ is a product state of four direction eigenstates relative to four classical directions $\mathbf{n}_i$, $\mathbf{m}_i$, $\mathbf{n}_j$, $\mathbf{m}_j$, we obtain
\begin{equation}\label{uncertainty principle probabilities classical}
    \begin{gathered}
        \Delta P_i\; \Delta P_j\geq \frac{(1-q)}{4} \Big|\qty(\mathbf{m}_i+ \mathbf{m}_j)\cdot \qty( \mathbf{n}_i\times \mathbf{n}_j) \\ -\qty( \mathbf{n}_i+ \mathbf{n}_j)\cdot \qty( \mathbf{m}_i\times \mathbf{m}_j)\Big|\;.
    \end{gathered}
\end{equation}
In \cref{appendix: geometrical examples}, we show that the coefficient multiplying $(1-q)$ lies within the interval $[0,4\sqrt{3}/9]$. In particular, the coefficient vanishes when either $(\mathbf{m}_i+\mathbf{m}_j)\parallel (\mathbf{n}_i+\mathbf{n}_j)$ or $(\mathbf{m}_i-\mathbf{m}_j+\mathbf{n}_i-\mathbf{n}_j)=\alpha\, (\mathbf{m}_i+\mathbf{m}_j) + \beta\,(\mathbf{n}_i+\mathbf{n}_j)$, where $\alpha,\beta$ are real coefficients compatible with the constraint that $\mathbf{m}_i,\mathbf{m}_j,\mathbf{n}_i,\mathbf{n}_j$ are unit vectors. For example, a physical configuration corresponding to this scenario occurs when all directions lie in the same plane. The conditions for obtaining $4\sqrt{3}/9$ as the coefficient multiplying $(1-q)$ are $\qty(\mathbf{n}_i-\mathbf{n}_j) \parallel \qty(\mathbf{m}_i-\mathbf{m}_j) \parallel \qty(\qty(\mathbf{n}_i+\mathbf{n}_j)\times\qty(\mathbf{m}_i+\mathbf{m}_j))$ and $\qty(\mathbf{n}_i+\mathbf{n}_j)\perp\qty(\mathbf{m}_i+\mathbf{m}_j)$, as proved in Appendix \ref{appendix: geometrical examples}. An explicit realisation of these conditions is
\begin{equation}
\begin{gathered}
    \mathbf{n}_i = \frac{1}{\sqrt{3}}(1,1,1)\;,\; \mathbf{n}_j = \frac{1}{\sqrt{3}}(-1,1,1)\;,\\ \mathbf{m}_i = \frac{1}{\sqrt{3}}(1,-1,1)\;\; \mathbf{m}_j = \frac{1}{\sqrt{3}}(-1,-1,1)\;,
\end{gathered}
\end{equation}
which can be visualised in appendix \ref{appendix: geometrical examples} in \cref{fig_app:sphere}. Of course, the expression for the uncertainty principle in \eqref{uncertainty principle probabilities} could also be computed on geometry states that are superpositions of directions eigenstates. This would result in a weighted sum of factors of the same form as the coefficient that appears on the right-hand side of \eqref{uncertainty principle probabilities classical}. Notice, however, that at this order of approximation it is impossible to appreciate interference effects from such superpositions, given that the commutator among probabilities \eqref{commutators probabilities} is diagonal in the direction operators.

At higher orders in $1-q$, the geometry states and direction operators are expected to display a substantially richer structure, to the point that the very notion of a sharp classical direction in space may cease to be meaningful. As we see here, at leading order, these complications are absent, and the above uncertainty relation admits a transparent interpretation in terms of the classical orientations of the measuring apparatuses.

Having derived the uncertainty principle for probability measurements \eqref{uncertainty principle probabilities} (and \eqref{uncertainty principle probabilities classical}), we understand that the probabilities operators in \eqref{probability operators braided} are much more non-classical than what we previously discussed in \cite{DEsposito:2024wru}. There, we argued that, being described by an operator, the probability of obtaining a certain outcome in a measurement is itself described by a probability distribution. In the present discussion, we find that there is a much richer structure which makes probabilities  indefinite in a quantum-mechanical fashion, in the sense that they cannot be assigned definite values simultaneously.  If one probability is sharply measured, then in general it is not possible to make any prediction about the outcome of the other probability measurement, and the order in which probabilities are measured matters. These features deepen the unpredictability of quantum theory, leveraging it to the level of probabilities themselves: just as the transition from classical mechanics to quantum mechanics makes individual measurement outcomes indefinite, in the same way the transition from quantum mechanics to doubly quantum mechanics makes even the probabilities of measurement outcomes fundamentally indefinite, in the sense that they cannot be assigned definite values simultaneously. 

Technically speaking, the uncertainty principle \eqref{uncertainty principle probabilities} is a direct consequence of the non-trivial commutation relations between distinct probability operators, given by \eqref{commutators probabilities}. At a structural level, the braiding construction developed in \cref{sec: braided DQM} implies that the Hilbert space $\mathcal{H}^A$ does not factorise into separate Hilbert spaces on which the different probability apparatuses act independently; rather, all measurement apparatuses share a single, common geometry state, on which the non-commuting probability operators act.

\begin{figure*}[t]
\centering
\begin{minipage}{0.48\textwidth}
    \centering
    \includegraphics[width=0.7\linewidth]{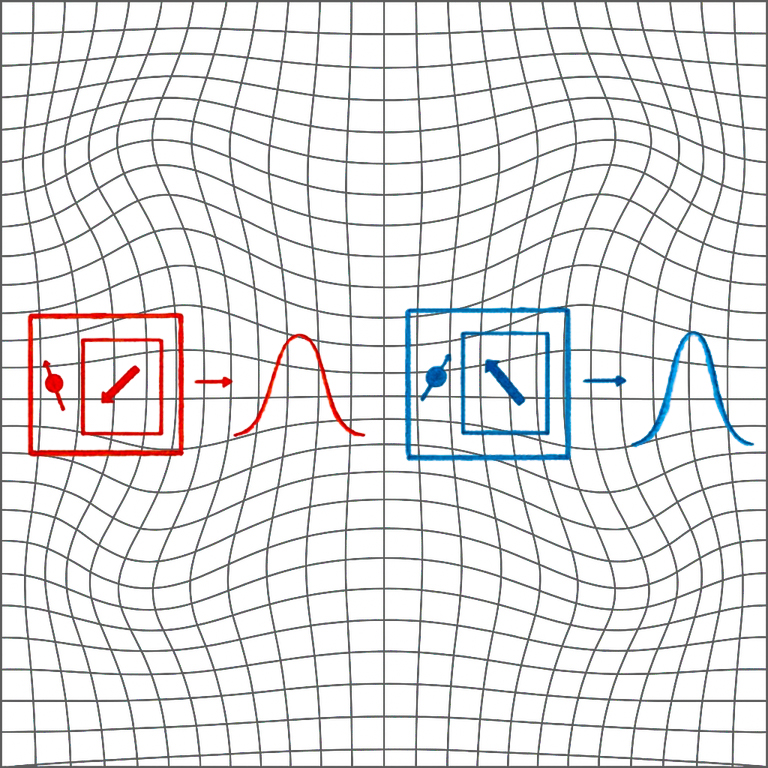}\\
    \vspace{7pt}
    \textbf{(a)}
\end{minipage}\hfill
\begin{minipage}{0.48\textwidth}
    \centering
    \includegraphics[width=0.7\linewidth]{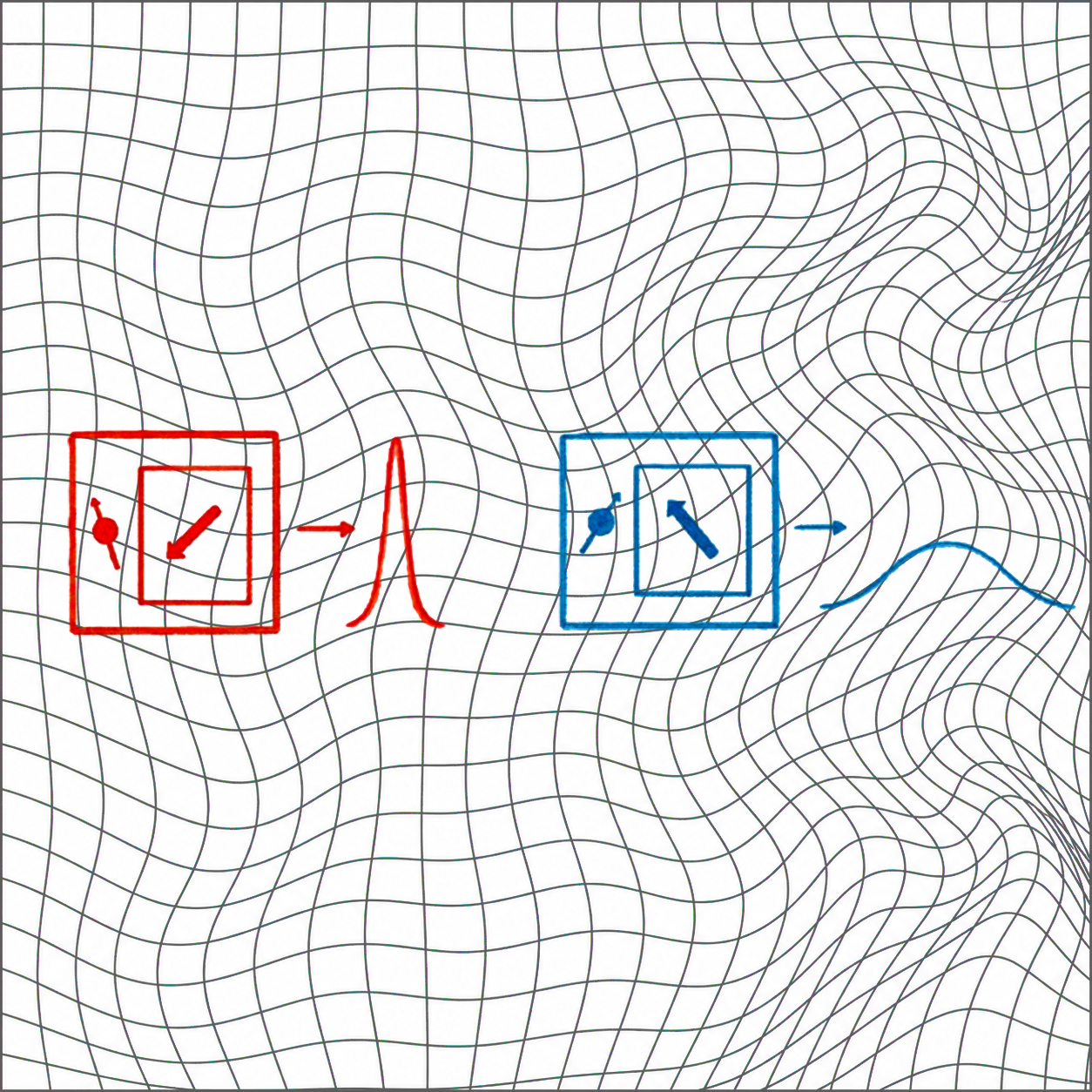}\\
    \vspace{7pt}
    \textbf{(b)}
\end{minipage}
\caption{A pictorial representation of the effect of probability measurements on geometry itself. On the left-hand side, the geometry states of the spin beams and Stern-Gerlach apparatuses are prepared in superpositions of probability eigenstates. These superpositions are represented by the Gaussians at the right of their respective probability measurement apparatus. The image on the right-hand side represents the situation once probability is measured with the red apparatus. The probability state collapses in a probability eigenstate, and consequently disturbs the state of the geometry which will be probed by a consequent probability measurement performed by the blue apparatus. This disturbance is pictorially represented by a more distorted spacetime grid on the side of the blue probability measurement apparatus. As an example, we have shown a situation in which the uncertainty in the probability measurement performed by the blue apparatus grows as a consequence of the probability measurement performed by the red apparatus.}
\label{fig:placeholder}
\end{figure*}

A more heuristic interpretation of this result can be offered if one takes at face value the idea that directions described by $SU_q(2)$ provide an effective description of quantum spacetime in some underlying theory of quantum gravity. In this picture, spacetime itself (or, in the present case, just spatial directions) is promoted to a genuine quantum degree of freedom, encoded in the states of the geometry. Once the standard collapse postulate of quantum mechanics is assumed, this quantum geometry is itself affected by the act of performing probability measurements: each measurement projects the underlying geometry state onto an eigenstate of the corresponding probability operator, thereby altering the very background against which subsequent measurements are interpreted. For a pictorial representation of this, see  Fig. \ref{fig:placeholder}. From this perspective, the incompatibility of probability measurements associated with non-commuting directions reflects the more fundamental fact that the geometry on which these probabilities are defined is itself quantum, and is therefore unavoidably disturbed by the measurement process.

\subsection{Communication protocol between Alice and Bob}

Now we return to the communication protocol between two observers who want to find out their relative orientation \cite{Amelino-Camelia:2022dsj, DEsposito:2024wru}. To do so, we specify \eqref{uncertainty principle probabilities classical} to the present case and write the states describing the experimental setup in the reference frame of $B$, who performs the measurements in the protocol. Doing so, the directions of the Stern-Gerlach devices used to perform the measurements coincide with the Cartesian axes of $B$, \emph{i.e.} $\mathbf{m}_i = \mathbf{i}$, with $i = x, y, z$. For what concerns the directions of the spins, they are described, in $B$'s reference frame by a rotation of the Cartesian axes of $A$, hence we denote them as $\mathbf{n}_i$, with $i=x,y,z$. 

The elements of the rotation matrix can then be written in terms of the probability operators as \cite{DEsposito:2024wru}
\begin{equation}
    R_{ij} \simeq 2\,P\qty(\mathbf{n}_i,\mathbf{j})-1\;,
\end{equation}
where we have neglected higher order corrections in $1-q$ since they will contribute to order $(1-q)^2$ to the uncertainty principle. We can immediately see that, in general, two rotation matrix elements do not commute and the measurements of two of them will be characterised by an uncertainty principle that at first order in $1-q$ reads
\begin{equation}\label{uncertainty principle matrix elements}
    \begin{gathered}
        \Delta R_{ij} \Delta R_{kl} \geq (1-q) \Big|\qty(\mathbf{j}+ \mathbf{l}\,)\cdot \qty( \mathbf{n}_i\times \mathbf{n}_k) \\ -\qty( \mathbf{n}_i+ \mathbf{n}_k)\cdot \qty( \mathbf{j}\times \mathbf{l}\,)\Big|\;.
    \end{gathered}
\end{equation}
This means that $B$ cannot sharply measure all the rotation matrix elements simultaneously, and the two observers cannot measure their relative orientations with arbitrary precision (see Appendix \ref{app3} for details). As an example, considering the case in which the reference frames of the two observers would be classically aligned ($R_{ij} = \delta_{ij}$), we have that $\mathbf{n}_i = \mathbf{i}$, therefore \eqref{uncertainty principle matrix elements} reads
\begin{equation}
    \Delta R_{ij} \Delta R_{kl} \geq (1-q)\abs{\epsilon_{jik} + \epsilon_{lik} - \epsilon_{ijl}-\epsilon_{kjl}}\;.
\end{equation}
From this we see, for instance, that $\Delta R_{xx} \Delta R_{yz} \geq 2(1-q)$, which reproduces the fact that even when observers would be aligned classically, they cannot sharply measure their relative orientation. This result is in agreement with the fact that, in our framework, directions associated with the messengers of the protocol and the measurement apparatuses are non-classical.

With this result, we come full circle. In fact, in the introduction we argued that non-classicality of spacetime could affect the very nature of probabilities in quantum measurements. We then derived an uncertainty relation between probabilities in a model of quantum group symmetries, and we now showed how this is reflected in quantum properties affecting the geometry of spacetime, in our model characterized by the relative orientation between observers.

\section{Conclusions}\label{sec: discussion}
In this letter, we have shown how properties of quantum spacetime naturally lead to a revision of our current understanding of quantum mechanics. For the first time, we derive a notion of indefinite probabilities, arising in spin measurements as a consequence of enforcing a deformed notion of rotation symmetry described by the quantum group $SU_q(2)$. The braiding structure of $SU_q(2)$ implies that probability measurements relative to different spin beams and Stern-Gerlach apparatuses are described by non-commuting probability operators, acting on geometry states characterising the orientations of the spins and the Stern-Gerlach devices themselves. The key result of our work, \cref{uncertainty principle probabilities}, is an uncertainty principle between consecutive probability measurements, which generally prevents them from being sharply measured simultaneously. Our results capture a deeper level of unpredictability than standard quantum theory, where indeterminacy is limited to measurement outcomes and does not affect the very probabilities of observing those outcomes. This non-commutativity of probabilities is also directly related to the non-classical structure of spacetime, as the rotation matrix elements defining the relative orientation between two observers become non-commutative too. Relative orientation hence acquires intrinsic quantum features.

The results and methodology presented in this letter pave the way for future investigations relating quantum spacetime properties to indefinite probabilities, extending the analysis beyond rotational symmetry to the full set of spacetime symmetries. The extension to other spacetime symmetries might also give new insights into the correspondence between quantum group structures and quantum reference frames, as suggested in \cite{Ballesteros:2025ypr}, where it was shown that a deformation of the Galilei symmetry group, at first order in the deformation parameter which is identified with the inverse of the mass of the reference frame, describes quantum reference frame transformations for ideal reference frames. It is also suggested that the full structure of the quantum group could describe transformations between non-ideal quantum reference frames. In this direction, the results of our work might be pivotal in establishing a connection between the $SU_q(2)$ quantum group and quantum reference frames for rotations introduced in \cite{Mikusch:2021kro}, as we already suggested in \cite{DEsposito:2024wru}. Such frames are defined by spin coherent states with some value $J$ of the spin, where a finite $J$ describes the case of non-ideal reference frames. By linking the deformation parameter $q$ to a function of such a spin variable, it might be possible to establish a correspondence between the quantum reference frames for rotations introduced in \cite{Mikusch:2021kro} and a quantum group deformation, much as what has been done for Galilean symmetries in \cite{Ballesteros:2025ypr}. The motivation for us to investigate this connection comes also from the ongoing work of Brukner and collaborators presented during conference talks \cite{slides,slides2}, which we became aware of in the advanced stages of our work. Indeed, within the context of quantum reference frames, they independently develop non-trivial commutation relations between relative frequencies in the limit of infinitely many repetitions.

Besides these possible connections with the quantum reference frames research
programme, we want to remark that there is a crucial difference with the case in which these deformations of symmetries originate from a quantum structure of spacetime. In these cases, such as what we consider in our work, the deformation of symmetries is of fundamental origin: it does not depend on the amount of resources or the precision of measurement devices. Therefore, the non-classical features and effects described by these deformations are intrinsic in the fabric of spacetime, and are unavoidable, even in ideal cases. This work then suggests that, in deepening our understanding of the nature of spacetime at the quantum level, we could be forced to extend the unpredictability of quantum theory, in such a way that even probabilities themselves cannot be described by definite values.

\begin{acknowledgments}

G.F. acknowledges financial support from the Blanceflor Foundation. D.F. acknowledges support from CSTQ: This work was made possible through the support of the WOST, WithOut SpaceTime project, supported by Grant ID\# 63683 from the John Templeton Foundation (JTF). The opinions expressed in this work are those of the authors and do not necessarily reflect the views of the John
Templeton Foundation. This work was carried out with the support of the Italian Ministry of University and Research (MUR) under the Funding for Research Projects (FIS 2), pursuant to Ministerial Decree No. 23314 of 11/12/2024 (Project Q-GraSp, FIS-2023-02629). This work contributes to the European Union COST Action CA23130 \emph{Bridging high and low energies in search of quantum gravity}.

\end{acknowledgments}

\providecommand{\noopsort}[1]{}\providecommand{\singleletter}[1]{#1}%

\providecommand{\href}[2]{#2}\begingroup\raggedright\endgroup

\newpage

\begin{onecolumngrid}

\appendix
\renewcommand{\theequation}{\thesection.\arabic{equation}}\setcounter{equation}{0}

\section{Details of the computations}\label{appendix: computations}

In this appendix we present some technical computations performed to derive the results of the main text.

\subsection{Details of the braiding relations}

As discussed in \cref{subsec: braiding}, the braiding relations are defined by

\begin{equation}\label{eq: braiding appendix}
\begin{gathered}
    u^\alpha \,w^\beta = q^{-1} R^{\alpha\beta}_{\gamma\delta}\, w^\gamma \, u^\delta\;, \\
    u^\alpha\, \bar{w}^\beta = R^{\alpha\beta}_{\gamma\delta}\, \bar{w}^\gamma \, u^\delta\;.
\end{gathered}
\end{equation}
where
\begin{equation}
    R^{\alpha\beta}_{\gamma\delta} = q\,\delta^\alpha_\gamma\,\delta^\beta_\delta + \epsilon^{\alpha\beta}\epsilon_{\gamma\delta}\;,\;\; \epsilon^{\alpha\beta} = \begin{pmatrix}
    0 & 1\\
    -q & 0 
\end{pmatrix}\;,\;\; \epsilon_{\alpha\beta} = \begin{pmatrix}
    0 & -q^{-1}\\
    1 & 0 
\end{pmatrix}\;.
\end{equation}
and $u^\alpha = \begin{pmatrix} x\\ y \end{pmatrix}$, $w^\alpha = \begin{pmatrix} a\\ c \end{pmatrix}$. With these definitions, \eqref{eq: braiding appendix} easily reads

\begin{equation}\label{eq: braiding commutation relation appendix}
    \begin{gathered}
        xa=ax\;,\quad qxc -cx=\qty(q-q^{-1})\,ay\;, \quad ay=qya\;,\quad yc=cy\\
        xc^*=qc^*x\;,\quad xa^*-a^*x=\qty(1-q^2)\,c^*y\;, \quad yc^*=c^*y\;, ya^*=qa^*y\;.
    \end{gathered}
\end{equation}
These equations, together with those pertaining the two copies $(x,y)$ and $(a,c)$, constitute the whole set of commutation relations satisfied by two braided copies of $SU_q(2)$. Notice that, if we take two identical $q$-spinors in \eqref{eq: braiding appendix}, we get exactly the commutation relations of the $SU_q(2)$ quantum group. This can be seen immediately by making the substitutions $a \to x$ and $c \to y$ (and analogously for $a^*$ and $c^*$) in \eqref{eq: braiding commutation relation appendix}, which amounts to taking two $u^\alpha$ $q$-spinors in \eqref{eq: braiding appendix}.

\subsection{All orders commutator}

In this appendix, we present the expression, at all-orders in $1-q$, for the commutator between two probability operators $P_i(\uparrow)$, $P_j(\uparrow)$ from which we explicitly derive the result \eqref{commutators probabilities}. 

The commutator between two probabilities reads
\begin{equation}
\begin{aligned}
\Big[P_i\qty(\uparrow)\,,\,P_j\qty(\uparrow)\Big] =\,&\frac{1}{q^3}(1 - q^2)\Big(
- q^2 a_j c_j^* a_i^* c_i
+ q^2 a_j c_j^* x_j^* y_j
- q a_j^* a_j c_i^* c_i
+ q^3 a_j^* a_j c_i^* c_i \\
&+ q a_j^* a_j y_j^* y_j
- q^3 a_j^* a_j y_j^* y_j
+ q^2 a_j^* c_j a_i c_i^*
- q^2 a_j^* c_j x_j y_j^* \\
&+ q^3 c_j^* c_j c_i^* c_i
- q^5 c_j^* c_j c_i^* c_i
+ q^4 a_j c_j^* a_i^* c_i y_j^* y_j
+ q^2 a_j c_j^* a_i^* c_i y_i^* y_i \\
&- q^2 a_j c_j^* c_i^* c_i x_j^* y_j
- a_j c_j^* c_i^* c_i x_i^* y_i
- q^2 a_j c_j^* x_j^* y_j y_i^* y_i
+ a_j c_j^* y_j^* y_j x_i^* y_i \\
&- q^2 a_j^* a_j a_i c_i^* x_i^* y_i
+ q^2 a_j^* a_j a_i^* c_i x_i y_i^*
+ q a_j^* a_j c_i^* c_i y_i^* y_i
- q^3 a_j^* a_j c_i^* c_i y_i^* y_i \\
&+ q^2 a_j^* a_j x_j y_j^* x_i^* y_i
- q^2 a_j^* a_j x_j^* y_j x_i y_i^*
- q a_j^* a_j y_j^* y_j y_i^* y_i
+ q^5 a_j^* a_j y_j^* y_j y_i^* y_i \\
&- q^4 a_j^* c_j a_i c_i^* y_j^* y_j
- q^2 a_j^* c_j a_i c_i^* y_i^* y_i
+ q^6 a_j^* c_j c_i^* c_i x_j y_j^*
+ q^4 a_j^* c_j c_i^* c_i x_i y_i^* \\
&+ q^2 a_j^* c_j x_j y_j^* y_i^* y_i
- q^4 a_j^* c_j y_j^* y_j x_i y_i^*
+ q^4 c_j^* c_j a_i c_i^* x_j^* y_j
- q^4 c_j^* c_j a_i^* c_i x_j y_j^* \\
&- q^3 c_j^* c_j c_i^* c_i y_j^* y_j
+ q^7 c_j^* c_j c_i^* c_i y_j^* y_j
- q^3 c_j^* c_j c_i^* c_i y_i^* y_i
+ q^5 c_j^* c_j c_i^* c_i y_i^* y_i
\Big)\,.
\end{aligned}
\end{equation}
By expanding the previous result at order $1-q$ we obtain
\begin{equation}\label{appendix1-q}
\begin{aligned}
\Big[P_i\qty(\uparrow)\,,\,P_j\qty(\uparrow)\Big] &\simeq\,2 (1 - q)\Big\{
y_i \, y_i^* \, x_j \, y_j^* \, a_i \, c_i
- y_i \, y_i^* \, x_j^* \, y_j \, a_i \, c_i^*
- x_j \, y_j^* \, a_j^* \, c_j
+ y_i \, y_i^* \, x_j \, y_j^* \, a_j^* \, c_j \\
&
+ a_i \, c_i^* \, a_j^* \, c_j
- y_i \, y_i^* \, a_i \, c_i^* \, a_j^* \, c_j
- y_j \, y_j^* \, a_i \, c_i^* \, a_j^* \, c_j
+ x_j \, y_j^* \, c_i \, c_i^* \, a_j^* \, c_j \\
&
+ x_j^* \, y_j \, a_j \, c_j^*
- y_i \, y_i^* \, x_j^* \, y_j \, a_j \, c_j^*
- a_i^* \, c_i \, a_j \, c_j^*
+ y_i \, y_i^* \, a_i^* \, c_i \, a_j \, c_j^* \\
&
+ y_j \, y_j^* \, a_i^* \, c_i \, a_j \, c_j^*
- x_j^* \, y_j \, c_i \, c_i^* \, a_j \, c_j^*
- x_j \, y_j^* \, a_i^* \, c_i \, c_j \, c_j^*
+ x_j^* \, y_j \, a_i \, c_i^* \, c_j \, c_j^* \\
&
+ x_i \, y_i^*
\Big[
(-y_j \, y_j^* + c_i \, c_i^*) a_j^* c_j
- a_i^* c_i (-1 + y_j y_j^* + c_j c_j^*) + x_j^* y_j (-1 + c_i c_i^* + c_j c_j^*)
\Big] \\
&
+ x_i^* \, y_i
\Big[
(y_j \, y_j^* - c_i \, c_i^*) a_j c_j^*
+ a_i c_i^* (-1 + y_j y_j^* + c_j c_j^*) - x_j y_j^* (-1 + c_i c_i^* + c_j c_j^*)
\Big]
\Big\}\;,\\
\end{aligned}
\end{equation}
where we can take all the operators appearing in \eqref{appendix1-q} as commuting, since  every non-commutative contribution would lead sub-leading terms
in $1-q$. Recalling our definition of quantum direction operators at order 0 in $1-q$ \eqref{directions}
\begin{equation}
    \begin{gathered}  \widehat{\mathbf{n}}_i \ebd \qty(x_i\, y^*_i + y_i\, x^*_i\,,\; \mathbbm{i}(y_i\, x_i^* - x_i\, y_i^*)\,,\; 1 - 2\,y_i \,y_i^*)\;,\\
\widehat{\mathbf{n}}_j \ebd 
\qty(
x_j\, y_j^* + y_j\, x_j^* \,,\;
\mathbbm{i}(y_j\, x_j^* - x_j\, y_j^*) \,,\;
1 - 2\, y_j\, y_j^*
)\;,\\  \widehat{\mathbf{m}}_i \ebd \qty(a_i\, c^*_i + c_i\, a^*_i\,,\; \mathbbm{i}(c_i\, a_i^* - a_i\, c_i^*)\,,\; 1 - 2\,c_i \,c_i^*)\;,\\
\widehat{\mathbf{m}}_j \ebd 
\qty(
a_j\, c_j^* + c_j\, a_j^* \,,\;
\mathbbm{i}(c_j\, a_j^* - a_j\, c_j^*) \,,\;
1 - 2\, c_j\, c_j^*
)\;,
    \end{gathered}
\end{equation}
we can rewrite \eqref{appendix1-q} as
\begin{equation}
\Big[P_i\qty(\uparrow)\,,\,P_j\qty(\uparrow)\Big] \simeq \frac{i}{2}(1-q)\Big\{\qty( \widehat{\mathbf{m}}_i+ \widehat{\mathbf{m}}_j)\cdot \qty( \widehat{\mathbf{n}}_i\times  \widehat{\mathbf{n}}_j) - \qty( \widehat{\mathbf{n}}_i+ \widehat{\mathbf{n}}_j)\cdot \qty( \widehat{\mathbf{m}}_i\times  \widehat{\mathbf{m}}_j)\Big\}\;,
\end{equation}
which coincides with \eqref{commutators probabilities}.

\subsection{Uncertainty principle for rotation matrix}\label{app3}
In this appendix, starting from \eqref{uncertainty principle matrix elements}, we prove that two observers, regardless of the rotation relating them, cannot sharply determine their relative orientation. 
The uncertainty principle between two generic entries of the rotation matrix reads
\begin{equation}\label{app: uncertR}
        \Delta R_{ij} \Delta R_{kl} \geq (1-q) \Big|\qty(\mathbf{j}+ \mathbf{l}\,)\cdot \qty( \mathbf{n}_i\times \mathbf{n}_k) -\qty( \mathbf{n}_i+ \mathbf{n}_k)\cdot \qty( \mathbf{j}\times \mathbf{l}\,)\Big|\;.
\end{equation}
We recall that the directions $\mathbf{n}_i$ are the results of the rotation of axes of $A$ with the rotation matrix $R$. Writing the generic direction $\mathbf{n}_i$ as $\mathbf{n}_i^a=R^{ab}i_b=R^{ab}\delta_{ib}=R^{ai}$, we can rewrite \eqref{app: uncertR} as
\begin{equation}\label{app: uncR2}
\Delta R_{ij}\,\Delta R_{kl}\ge (1-q)\,
\Big|
\tensor{\varepsilon}{_j}{^b}{^c}R_{bi}R_{ck}
+
\tensor{\varepsilon}{_l}{^b}{^c}R_{bi}R_{ck}
-
\tensor{\varepsilon}{^a}{_j}{_l}\left(R_{ai}+R_{ak}\right)
\Big|\;.
\end{equation}
Using the fact that
\begin{equation}
\varepsilon^{j\beta\gamma} R_{\beta i} R_{\gamma k}
=
\varepsilon_{ikm} R_{jm}\;,
\end{equation}
we can rewrite \eqref{app: uncR2} as
\begin{equation}
\Delta R_{ij}\,\Delta R_{kl}\geq (1-q) \Big|
\tensor{\varepsilon}{_i}{_k}{^m}(R_{jm}+R_{lm})
-
\tensor{\varepsilon}{^a}{_j}{_l}(R_{ak}+R_{ai})\Big|\;.
\end{equation}
Let us denote the term inside the absolute value by $E_{ijkl}$. For a rotation matrix to be sharp, we should have $E_{ijkl} = 0$, $\forall\, i,j,k,l$. In this case, choosing some combinations of indexes, we have
\begin{equation}
\begin{gathered}
     E_{yxzx} - E_{xyxx} = 2\qty(R_{xx} - R_{zx}) = 0\;,\\
     E_{xxzx} - E_{yxyy} = -2\qty(R_{xy} - R_{zy}) = 0\;,\\
     E_{xxyx} - E_{zyzx} = 2\qty(R_{xz} - R_{zz}) = 0\;.
\end{gathered}
\end{equation}
This set of equations implies that the first and third rows of $R$ are equal, meaning that $\det(R) = 0$, which is not possible since $R \in SO(3)$. Therefore, $E_{ijkl}$ cannot be zero for every possible choice of indexes, and this means that there cannot be a sharp rotation matrix.

\section{Geometrical examples for the commutator}\label{appendix: geometrical examples}

The uncertainty principle \eqref{uncertainty principle probabilities classical} can be rewritten in terms of the differences and sums of the unit vectors involved as
\begin{equation}\label{eq_app: uncertainty principle probabilities classical}
    \Delta P_i\; \Delta P_j\geq \frac{(1-q)}{8} \Big|\qty(\mathbf{D}_n+ \mathbf{D}_m)\cdot \qty( \mathbf{S}_n\times \mathbf{S}_m)\Big|\;,
\end{equation}
where $\mathbf{S}_n = \mathbf{n}_i+\mathbf{n}_j$, $\mathbf{D}_n = \mathbf{n}_i - \mathbf{n}_j$. The conditions of minimum uncertainty are easily derived from this. Indeed, either the sums are parallel, $\mathbf{S}_n \parallel \mathbf{S}_m$ or $\mathbf{D}_n+ \mathbf{D}_m$ is perpendicular to $\mathbf{S}_n\times \mathbf{S}_m$, which means that $\mathbf{D}_n+ \mathbf{D}_m$ lies the plane defined by the vectors $\mathbf{S}_n$ and $\mathbf{S}_m$ and is therefore a linear combination of these two vector
\begin{equation}\label{eq: app planar condition}
    \mathbf{D}_n+ \mathbf{D}_m = \alpha\, \mathbf{S}_n + \beta\, \mathbf{S}_m\;,
\end{equation}
where $\alpha$ and $\beta$ are real coefficients that should satisfy a compatibility constraint with the condition that $\mathbf{n}_i, \mathbf{n}_j, \mathbf{m}_i, \mathbf{m}_j$ are unit vectors. In particular, from this condition we have
\begin{equation}
    \mathbf{S}_n \cdot \mathbf{D}_n = \|\mathbf{n}_i\|^2 - \|\mathbf{n}_j\|^2 = 1 - 1 = 0\;,
\end{equation}
and similarly, $\mathbf{S}_m \cdot \mathbf{D}_m = 0$. Furthermore, the squared norms satisfy $\|\mathbf{D}\|^2 = 4 - \|\mathbf{S}\|^2$. We can rewrite \eqref{eq: app planar condition} as $\mathbf{D}_n - \alpha \mathbf{S}_n = \beta \mathbf{S}_m - \mathbf{D}_m$ and, squaring both sides, from the orthogonality condition we have $ \|\mathbf{D}_n\|^2 + \alpha^2\|\mathbf{S}_n\|^2 = \beta^2\|\mathbf{S}_m\|^2 + \|\mathbf{D}_m\|^2$. Substituting $\|\mathbf{D}\|^2 = 4 - \|\mathbf{S}\|^2$ we finally arrive at the constraint that the coefficients $\alpha$ and $\beta$ should satisfy
\begin{equation}
    (\alpha^2 - 1)\|\mathbf{S}_n\|^2 = (\beta^2 - 1)\|\mathbf{S}_m\|^2\;.
\end{equation}
Notice that \eqref{eq: app planar condition} also includes the situation in which $\mathbf{D}_n = - \mathbf{D}_m$.

The maximum value of the coefficient appearing in \eqref{eq_app: uncertainty principle probabilities classical} is derived as follows. Calling the relative angles between the two $\mathbf{n}$ vectors and the two $\mathbf{m}$ vectors $\theta_n$ and $\theta_m$ respectively, we have that
\begin{equation}
    \Big|\qty(\mathbf{D}_n+ \mathbf{D}_m)\cdot \qty( \mathbf{S}_n\times \mathbf{S}_m)\Big| \leq 8\qty(\sin{\frac{\theta_n}{2}}+\sin{\frac{\theta_m}{2}})\cos{\frac{\theta_n}{2}}\cos{\frac{\theta_m}{2}}\sin \phi\;,
\end{equation}
where $\phi$ is the angle between $\mathbf{S}_m$ and $\mathbf{S}_n$. The equality in the last equation, so the maximum value of the coefficients in \eqref{eq_app: uncertainty principle probabilities classical}, is reached when $\mathbf{D}_n \parallel \mathbf{D}_m \parallel (\mathbf{S}_n\times\mathbf{S}_m)$, which can be realised when $\mathbf{D}_n \parallel \mathbf{D}_m$ are along the same axis, perpendicular to the plane defined by $\mathbf{S}_n$ and $\mathbf{S}_m$. The degree of freedom $\phi$ is therefore the angle between these two vectors in this plane, and therefore the maximum value of the coefficient in \eqref{eq_app: uncertainty principle probabilities classical} is reached when these two vectors are orthogonal in this plane, $\mathbf{S}_n \perp \mathbf{S}_m$. Using $\sin{\frac{\theta_n}{2}} = x$ and $\sin{\frac{\theta_m}{2}} = y$, we can then maximise the resulting function of $x,y$ by computing derivatives with respect to these variables and find that the maximum is reached when $x=y=\frac{1}{\sqrt{3}}$, which yields
\begin{equation}\label{eq_app: max_uncertainty}
    \qty(\Delta P_i\; \Delta P_j)^{\text{max}} \geq \frac{4\sqrt{3}}{9}(1-q)\;.
\end{equation}
A specific example of a configuration in which this maximum value for the lower bound of the uncertainty is reached is when $\mathbf{n}_i = \frac{1}{\sqrt{3}}(1,1,1)$, $\mathbf{n}_j = \frac{1}{\sqrt{3}}(-1,1,1)$, $\mathbf{m}_i = \frac{1}{\sqrt{3}}(1,-1,1)$ and $\mathbf{m}_j = \frac{1}{\sqrt{3}}(-1,-1,1)$. Geometrically, this is depicted in \cref{fig_app:sphere}.
\begin{center}
    \begin{figure}[]
    \includegraphics[scale=0.45]{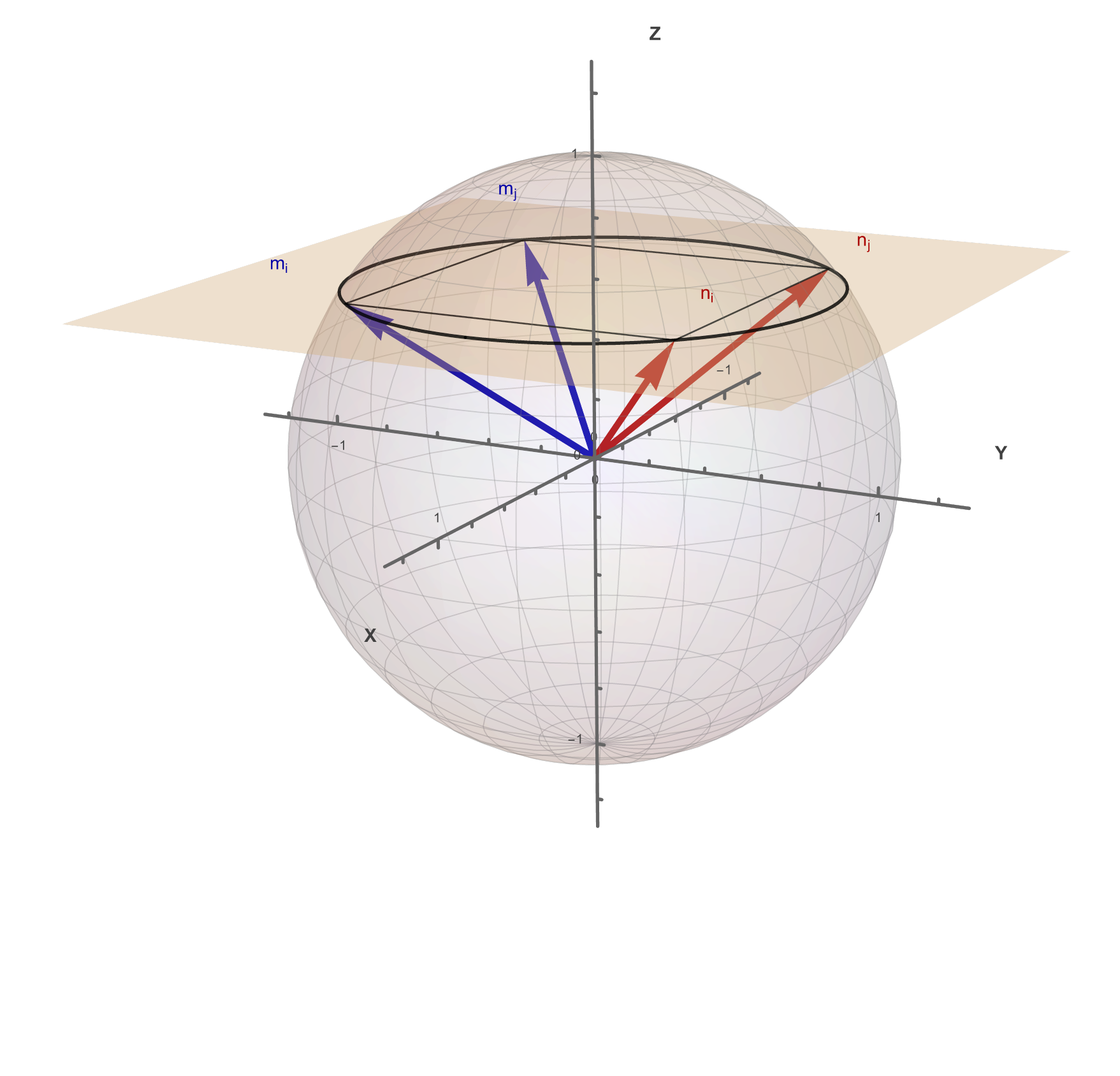}
    \caption{One possible geometrical realisation of \eqref{eq_app: max_uncertainty}. The four vectors $\mathbf{n}_i = \frac{1}{\sqrt{3}}(1,1,1)$, $\mathbf{n}_j = \frac{1}{\sqrt{3}}(-1,1,1)$, $\mathbf{m}_i = \frac{1}{\sqrt{3}}(1,-1,1)$ and $\mathbf{m}_j = \frac{1}{\sqrt{3}}(-1,-1,1)$ point from the origin to the vertices of the square inscribed in the circumference formed by the intersection of the sphere of radius $1$, with centre in the origin, and the plane $z=\frac{1}{\sqrt{3}}$.}
    \label{fig_app:sphere}
\end{figure}
\end{center}

\end{onecolumngrid}


\begin{thebibliography}{10}

\bibitem{Bartlett:2006tzx}
S.D.~Bartlett, T.~Rudolph and R.W.~Spekkens, \emph{{Reference frames,
  superselection rules, and quantum information}},
  \href{https://doi.org/10.1103/RevModPhys.79.555}{\emph{Rev. Mod. Phys.}
  {\bfseries 79} (2007) 555}
  [\href{https://arxiv.org/abs/quant-ph/0610030}{{\ttfamily
  quant-ph/0610030}}].

\bibitem{Amelino-Camelia:2022dsj}
G.~Amelino-Camelia, V.~D'Esposito, G.~Fabiano, D.~Frattulillo, P.A.~Höhn and
  F.~Mercati, \emph{{Quantum Euler angles and agency-dependent space-time}},
  \href{https://doi.org/10.1093/ptep/ptae015}{\emph{PTEP} {\bfseries 2024}
  (2024) 033A01} [\href{https://arxiv.org/abs/2211.11347}{{\ttfamily
  2211.11347}}].

\bibitem{DEsposito:2024wru}
V.~D'Esposito, G.~Fabiano, D.~Frattulillo and F.~Mercati, \emph{{Doubly Quantum
  Mechanics}}, \href{https://doi.org/10.22331/q-2025-04-24-1721}{\emph{Quantum}
  {\bfseries 9} (2025) 1721}
  [\href{https://arxiv.org/abs/2412.05997}{{\ttfamily 2412.05997}}].

\bibitem{Rovelli:1994ge}
C.~Rovelli and L.~Smolin, \emph{{Discreteness of area and volume in quantum
  gravity}}, \href{https://doi.org/10.1016/0550-3213(95)00150-Q}{\emph{Nucl.
  Phys. B} {\bfseries 442} (1995) 593}
  [\href{https://arxiv.org/abs/gr-qc/9411005}{{\ttfamily gr-qc/9411005}}].

\bibitem{Ashtekar:1996eg}
A.~Ashtekar and J.~Lewandowski, \emph{{Quantum theory of geometry. 1: Area
  operators}}, \href{https://doi.org/10.1088/0264-9381/14/1A/006}{\emph{Class.
  Quant. Grav.} {\bfseries 14} (1997) A55}
  [\href{https://arxiv.org/abs/gr-qc/9602046}{{\ttfamily gr-qc/9602046}}].

\bibitem{Freidel:2002hx}
L.~Freidel, E.R.~Livine and C.~Rovelli, \emph{{Spectra of length and area in
  (2+1) Lorentzian loop quantum gravity}},
  \href{https://doi.org/10.1088/0264-9381/20/8/304}{\emph{Class. Quant. Grav.}
  {\bfseries 20} (2003) 1463}
  [\href{https://arxiv.org/abs/gr-qc/0212077}{{\ttfamily gr-qc/0212077}}].

\bibitem{Amelino-Camelia:2008fcv}
G.~Amelino-Camelia, G.~Gubitosi and F.~Mercati, \emph{{Discreteness of area in
  noncommutative space}},
  \href{https://doi.org/10.1016/j.physletb.2009.04.045}{\emph{Phys. Lett. B}
  {\bfseries 676} (2009) 180}
  [\href{https://arxiv.org/abs/0812.3663}{{\ttfamily 0812.3663}}].

\bibitem{Majid:2006xn}
S.~Majid, \emph{{Algebraic approach to quantum gravity. II: Noncommutative
  spacetime}},  [\href{https://arxiv.org/abs/hep-th/0604130}{{\ttfamily
  hep-th/0604130}}].

\bibitem{Lizzi:2018qaf}
F.~Lizzi, M.~Manfredonia, F.~Mercati and T.~Poulain, \emph{{Localization and
  Reference Frames in $\kappa$-Minkowski Spacetime}},
  \href{https://doi.org/10.1103/PhysRevD.99.085003}{\emph{Phys. Rev. D}
  {\bfseries 99} (2019) 085003}
  [\href{https://arxiv.org/abs/1811.08409}{{\ttfamily 1811.08409}}].

\bibitem{Majid:2026ptz}
S.~Majid, \emph{{Algebraic approach to quantum gravity IV: applications}},
  [\href{https://arxiv.org/abs/2604.06118}{{\ttfamily 2604.06118}}].

\bibitem{Snyder:1947}
H.S.~Snyder, \emph{Quantized space-time},
  \href{https://doi.org/10.1103/PhysRev.71.38}{\emph{Phys. Rev.} {\bfseries 71}
  (1947) 38}.

\bibitem{Seiberg:1999vs}
N.~Seiberg and E.~Witten, \emph{{String theory and noncommutative geometry}},
  \href{https://doi.org/10.1088/1126-6708/1999/09/032}{\emph{JHEP} {\bfseries
  09} (1999) 032} [\href{https://arxiv.org/abs/hep-th/9908142}{{\ttfamily
  hep-th/9908142}}].

\bibitem{Douglas:2001ba}
M.R.~Douglas and N.A.~Nekrasov, \emph{{Noncommutative field theory}},
  \href{https://doi.org/10.1103/RevModPhys.73.977}{\emph{Rev. Mod. Phys.}
  {\bfseries 73} (2001) 977}
  [\href{https://arxiv.org/abs/hep-th/0106048}{{\ttfamily hep-th/0106048}}].

\bibitem{Brandenberger:2002nq}
R.~Brandenberger and P.-M.~Ho, \emph{{Noncommutative space-time, stringy
  space-time uncertainty principle, and density fluctuations}},
  \href{https://doi.org/10.1103/PhysRevD.66.023517}{\emph{Phys. Rev. D}
  {\bfseries 66} (2002) 023517}
  [\href{https://arxiv.org/abs/hep-th/0203119}{{\ttfamily hep-th/0203119}}].

\bibitem{Freidel:2005me}
L.~Freidel and E.R.~Livine, \emph{{3D Quantum Gravity and Effective
  Noncommutative Quantum Field Theory}},
  \href{https://doi.org/10.1103/PhysRevLett.96.221301}{\emph{Phys. Rev. Lett.}
  {\bfseries 96} (2006) 221301}
  [\href{https://arxiv.org/abs/hep-th/0512113}{{\ttfamily hep-th/0512113}}].

\bibitem{Oriti:2009pb}
D.~Oriti, \emph{{Emergent non-commutative matter fields from Group Field Theory
  models of quantum spacetime}},
  \href{https://doi.org/10.1088/1742-6596/174/1/012047}{\emph{J. Phys. Conf.
  Ser.} {\bfseries 174} (2009) 012047}
  [\href{https://arxiv.org/abs/0903.3970}{{\ttfamily 0903.3970}}].

\bibitem{Amelino-Camelia:2016gfx}
G.~Amelino-Camelia, M.M.~da~Silva, M.~Ronco, L.~Cesarini and O.M.~Lecian,
  \emph{{Spacetime-noncommutativity regime of Loop Quantum Gravity}},
  \href{https://doi.org/10.1103/PhysRevD.95.024028}{\emph{Phys. Rev. D}
  {\bfseries 95} (2017) 024028}
  [\href{https://arxiv.org/abs/1605.00497}{{\ttfamily 1605.00497}}].

\bibitem{Szabo:2001kg}
R.J.~Szabo, \emph{{Quantum field theory on noncommutative spaces}},
  \href{https://doi.org/10.1016/S0370-1573(03)00059-0}{\emph{Phys. Rept.}
  {\bfseries 378} (2003) 207}
  [\href{https://arxiv.org/abs/hep-th/0109162}{{\ttfamily hep-th/0109162}}].

\bibitem{Amelino-Camelia:2008aez}
G.~Amelino-Camelia, \emph{{Quantum-Spacetime Phenomenology}},
  \href{https://doi.org/10.12942/lrr-2013-5}{\emph{Living Rev. Rel.} {\bfseries
  16} (2013) 5} [\href{https://arxiv.org/abs/0806.0339}{{\ttfamily
  0806.0339}}].

\bibitem{majid_1995}
S.~Majid, \emph{Foundations of Quantum Group Theory}, Cambridge University
  Press (1995),
  \href{https://doi.org/10.1017/CBO9780511613104}{10.1017/CBO9780511613104}.

\bibitem{Arzano:2022nlo}
M.~Arzano, V.~D'Esposito and G.~Gubitosi, \emph{{Fundamental decoherence from
  quantum spacetime}},
  \href{https://doi.org/10.1038/s42005-023-01159-3}{\emph{Commun. Phys.}
  {\bfseries 6} (2023) 242} [\href{https://arxiv.org/abs/2208.14119}{{\ttfamily
  2208.14119}}].

\bibitem{Ballesteros:2025ypr}
A.~Ballesteros, D.~Fernandez-Silvestre, F.~Giacomini and G.~Gubitosi,
  \emph{{Quantum Galilei group as quantum reference frame transformations}},
  [\href{https://arxiv.org/abs/2504.00569}{{\ttfamily 2504.00569}}].

\bibitem{Gubitosi:2021itz}
G.~Gubitosi, F.~Lizzi, J.J.~Relancio and P.~Vitale, \emph{{Double
  quantization}},
  \href{https://doi.org/10.1103/PhysRevD.105.126013}{\emph{Phys. Rev. D}
  {\bfseries 105} (2022) 126013}
  [\href{https://arxiv.org/abs/2112.11401}{{\ttfamily 2112.11401}}].

\bibitem{Arzano:2026gng}
M.~Arzano, A.~Del~Prete and D.~Frattulillo, \emph{{Quantum Evolution of Hopf
  Algebra Hamiltonians}},  [\href{https://arxiv.org/abs/2602.07887}{{\ttfamily
  2602.07887}}].

\bibitem{Arzano:2026fbz}
M.~Arzano, G.~Chirco and J.~Kowalski-Glikman, \emph{{Bias in Local Spin
  Measurements from Deformed Symmetries}},
  [\href{https://arxiv.org/abs/2603.08618}{{\ttfamily 2603.08618}}].

\bibitem{Giacomini:2017zju}
F.~Giacomini, E.~Castro-Ruiz and v.~Brukner, \emph{{Quantum mechanics and the
  covariance of physical laws in quantum reference frames}},
  \href{https://doi.org/10.1038/s41467-018-08155-0}{\emph{Nature Commun.}
  {\bfseries 10} (2019) 494}
  [\href{https://arxiv.org/abs/1712.07207}{{\ttfamily 1712.07207}}].

\bibitem{Woronowicz:1987vs}
S.L.~Woronowicz, \emph{{Compact matrix pseudogroups}},
  \href{https://doi.org/10.1007/BF01219077}{\emph{Commun. Math. Phys.}
  {\bfseries 111} (1987) 613}.

\bibitem{Major:1995yz}
S.~Major and L.~Smolin, \emph{{Quantum deformation of quantum gravity}},
  \href{https://doi.org/10.1016/0550-3213(96)00259-3}{\emph{Nucl. Phys. B}
  {\bfseries 473} (1996) 267}
  [\href{https://arxiv.org/abs/gr-qc/9512020}{{\ttfamily gr-qc/9512020}}].

\bibitem{Freidel:1998pt}
L.~Freidel and K.~Krasnov, \emph{{Spin foam models and the classical action
  principle}}, \href{https://doi.org/10.4310/ATMP.1998.v2.n6.a1}{\emph{Adv.
  Theor. Math. Phys.} {\bfseries 2} (1999) 1183}
  [\href{https://arxiv.org/abs/hep-th/9807092}{{\ttfamily hep-th/9807092}}].

\bibitem{Girelli:2022foc}
F.~Girelli and M.~Laudonio, \emph{{Group field theory on quantum groups}},
  [\href{https://arxiv.org/abs/2205.13312}{{\ttfamily 2205.13312}}].

\bibitem{Bianchi:2011uq}
E.~Bianchi and C.~Rovelli, \emph{{Note on the geometrical interpretation of
  quantum groups and non-commutative spaces in gravity}},
  \href{https://doi.org/10.1103/PhysRevD.84.027502}{\emph{Phys. Rev. D}
  {\bfseries 84} (2011) 027502}
  [\href{https://arxiv.org/abs/1105.1898}{{\ttfamily 1105.1898}}].

\bibitem{hoehn2019trinity}
P.A.~H{\"o}hn, A.R.~Smith and M.P.~Lock, \emph{Trinity of relational quantum
  dynamics}, \href{https://doi.org/10.1103/PhysRevD.104.066001}{\emph{Phys.
  Rev. D} {\bfseries 104} (2021) 066001}
  [\href{https://arxiv.org/abs/1912.00033}{{\ttfamily 1912.00033}}].

\bibitem{giacomini2019relativistic}
F.~Giacomini, E.~Castro-Ruiz and {\v{C}}.~Brukner, \emph{Relativistic quantum
  reference frames: the operational meaning of spin},
  \href{https://doi.org/10.1103/PhysRevLett.123.090404}{\emph{Phys. Rev. Lett.}
  {\bfseries 123} (2019) 090404}
  [\href{https://arxiv.org/abs/1811.08228}{{\ttfamily 1811.08228}}].

\bibitem{streiter2020relativistic}
L.F.~Streiter, F.~Giacomini and {\v{C}}.~Brukner, \emph{{Relativistic Bell Test
  within Quantum Reference Frames}},
  \href{https://doi.org/10.1103/PhysRevLett.126.230403}{\emph{Phys. Rev. Lett.}
  {\bfseries 126} (2021) 230403}
  [\href{https://arxiv.org/abs/2008.03317}{{\ttfamily 2008.03317}}].

\bibitem{de2020quantum}
A.-C.~de~la Hamette and T.D.~Galley, \emph{Quantum reference frames for general
  symmetry groups},
  \href{https://doi.org/10.22331/q-2020-11-30-367}{\emph{{Quantum}} {\bfseries
  4} (2020) 367} [\href{https://arxiv.org/abs/2004.14292}{{\ttfamily
  2004.14292}}].

\bibitem{Mikusch:2021kro}
M.~Mikusch, L.C.~Barbado and v.~Brukner, \emph{{Transformation of spin in
  quantum reference frames}},
  \href{https://doi.org/10.1103/PhysRevResearch.3.043138}{\emph{Phys. Rev.
  Res.} {\bfseries 3} (2021) 043138}
  [\href{https://arxiv.org/abs/2103.05022}{{\ttfamily 2103.05022}}].

\bibitem{hoehn2021quantum}
S.~Ahmad~Ali, T.D.~Galley, P.A.~H{\"o}hn, M.P.E.~Lock and A.R.H.~Smith,
  \emph{{Quantum relativity of subsystems}},
  \href{https://doi.org/10.1103/PhysRevLett.128.170401}{\emph{Phys. Rev. Lett.}
  {\bfseries 128} (2022) 170401}
  [\href{https://arxiv.org/abs/2103.01232}{{\ttfamily 2103.01232}}].

\bibitem{de2021entanglement}
A.-C.~de~la Hamette, S.L.~Ludescher and M.P.~Mueller,
  \emph{{Entanglement-Asymmetry Correspondence for Internal Quantum Reference
  Frames}}, \href{https://doi.org/10.1103/PhysRevLett.129.260404}{\emph{Phys.
  Rev. Lett.} {\bfseries 129} (2022) 260404}
  [\href{https://arxiv.org/abs/2112.00046}{{\ttfamily 2112.00046}}].

\bibitem{giacomini2021spacetime}
F.~Giacomini, \emph{{Spacetime Quantum Reference Frames and superpositions of
  proper times}},
  \href{https://doi.org/10.22331/q-2021-07-22-508}{\emph{Quantum} {\bfseries 5}
  (2021) 508} [\href{https://arxiv.org/abs/2101.11628}{{\ttfamily
  2101.11628}}].

\bibitem{Cepollaro2024}
C.~Cepollaro, A.~Akil, P.~Cieśliński, A.-C.~de~la Hamette and
  {\v{C}}.~Brukner, \emph{The sum of entanglement and subsystem coherence is
  invariant under quantum reference frame transformations},
  \href{https://doi.org/10.1103/h6b3-y4vt}{\emph{Phys. Rev. Lett.} {\bfseries
  135} (2025) 010201} [\href{https://arxiv.org/abs/2406.19448}{{\ttfamily
  2406.19448}}].

\bibitem{Garmier2025}
S.C.~Garmier, L.~Hausmann and E.~Castro-Ruiz, \emph{{The Perspectives of
  Non-Ideal Quantum Reference Frames}},
  [\href{https://arxiv.org/abs/2512.19343}{{\ttfamily 2512.19343}}].

\bibitem{Xing-ChangSong_1992}
X.-C.~Song, \emph{Spinor analysis for quantum group suq(2)},
  \href{https://doi.org/10.1088/0305-4470/25/10/021}{\emph{Journal of Physics
  A: Mathematical and General} {\bfseries 25} (1992) 2929}.

\bibitem{slides}
E.~Castro-Ruiz, N.~Cohen, L.~Barbado and {\v{C}}.~Brukner, \emph{{Indefinite
  Probabilities from finite quantum reference frames}},
  {\emph{\href{https://www.oist.jp/conference/qrf-2025\#Brukner}{QRF 2025,
  Okinawa}} ($16^{\text{th}}$ October\,\, 2025) }.

\bibitem{slides2}
E.~Castro-Ruiz, N.~Cohen, L.~Barbado and {\v{C}}.~Brukner, \emph{{Quantum
  Incompatibility of Relative Frequencies}},
  {\emph{\href{https://quantum.physics.sk/ocqg2026/src/schedule.html}{Observers
  and Causality in Quantum Gravity, Bratislava}} ($23^{\text{rd}}$ April\,\,
  2026) }.

\end{thebibliography}
\end{document}